\documentclass[aps,physrev,twocolumn,superscriptaddress]{revtex4-1}
\usepackage{epsfig,graphicx,amssymb,color,amssymb,amsmath,mathrsfs,bm}
\newcommand{\dd}{\overset{\text{ \tiny $\leftrightarrow$}}}

\begin{document}
\title{Non-equilibrium states of a plasmonic Dicke model with coherent and dissipative surface plasmon-quantum emitter interactions}

\author{Andrei Piryatinski}
\email[]{apiryat@lanl.gov}
\affiliation{Theoretical Division, Los Alamos National Laboratory, Los Alamos, NM 87545 }

\author{Oleksiy Roslyak}
\affiliation{Department of Physics and Engineering Physics, Fordham University, Bronx, New York 10458, USA}

\author{Hao Li}
\affiliation{Department of Chemistry, University of Houston, Houston, TX 77204}

\author{Eric R. Bittner}
\email[]{ebittner@central.uh.edu}
\affiliation{Department of Chemistry, University of Houston, Houston, TX 77204}

\date{\today}

\begin{abstract}
Hybrid photonic-plasmonic nanostructures allow one to engineer coupling of quantum emitters and cavity modes accounting for the direct coherent and environment mediated dissipative pathways. Using generalized plasmonic Dicke model, we explore the non-equilibrium phase diagram with respect to these interactions. The analysis shows that their interplay results in the extension of the superradiant and regular lasing states to the dissipative coupling regime and an emergent lasing phase without population inversion having boundary with the superradiant and normal states. Calculated photon emission spectra are demonstrated to carry distinct signatures of these phases.  
\end{abstract}
\maketitle


\section{Introduction}

In quantum plasmonics, highly polarizable metal nanostructures supporting surface plasmon modes provide a source of strong enhancement in the photon local density of states, an effect similar to a low-Q optical cavity~\cite{TameNatPhys:2013}. Technological flexibility in the design of plasmonic cavities allows one to engineer surface plasmon states and their interactions with quantum emitters (QE), e.g., fluorescent dyes or semiconductor nanostructures, leading to potentially desirable cooperative properties~\cite{TameNatPhys:2013,HollingsworthMRSB:2015,SukharevJPCM:2017}.  The strong (ultra strong) coupling regimes, when the surface plasmon-QE interaction strength exceeds the total cavity losses (becomes comparable with the QE energy), opens new opportunities for non-equilibrium exciton-plasmon polariton condensation, non-linear emission, and lasing~\cite{RamezaniJOSAB:2019,KockumNatRevPhys:2019}.  

In many cases, the experimental demonstration of the effect was preceded by theoretical analysis. 
 For instance, theoretical studies of surface plasmon-induced superradiant and sub-radiant Dicke states reveal how their frequency and time-domain emission features depend on the cavity geometry, composition, and environment fluctuations~\cite{Pustovit:10,TemnovPRL:2005}. Rapid progress in the development of nanoscale surface plasmon laser, often referred to as the ``spaser", has been reported~\cite{StockmanJOpt:2010,BeriniNatPh:2012,WeiZhouNNano:2013,RichterPRB:2015,ZhangJCP:2015}. Theoretical analysis of critical phenomena such as Bose-Einstein condensation (BEC) of the surface plasmon-exciton-polaritons in plasmonic lattices and arrays~\cite{MartikainenPRA:2014,ZasterJPCS:2016} were followed by reports claiming experimental observation of a thermalized room temperature and non-equilibrium BEC along with polariton lasing~\cite{HakalaNP:2018,RamezaniJOSAB:2019}.

\begin{figure}[b]
\includegraphics[width=0.5\textwidth]{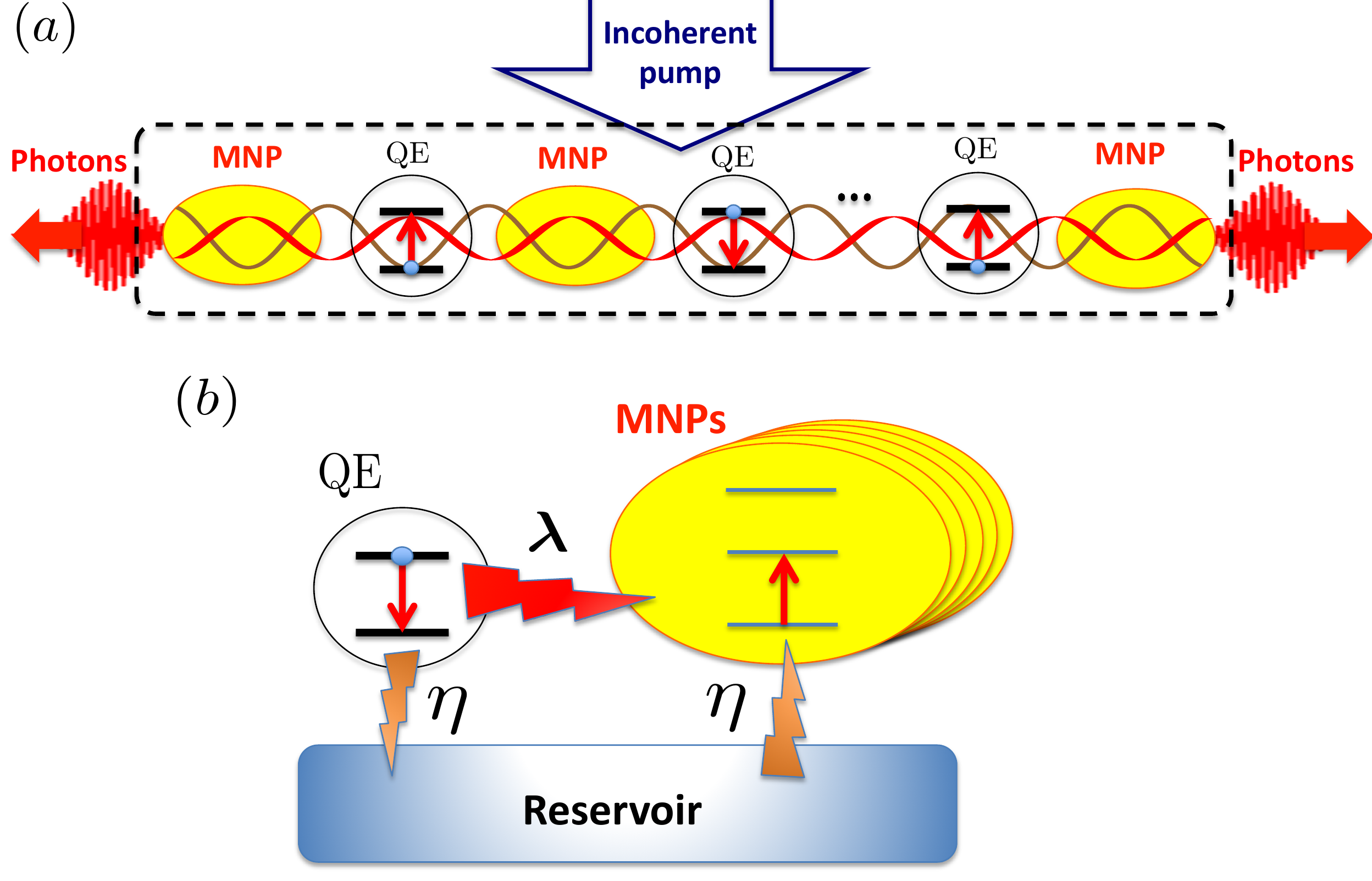}
\caption{\color{black}(a) An example of plasmonic cavity filled with QEs coupled to the SPCM (red wavy line) and dark plasmon mode (brown wavy line) both due to an array of metal nanoparticles (MNPs). The SPCM also interacts with the dark plasmon mode. The QEs are subject to an incoherent pumping and the SPCM is the source of photons emitted outside the cavity. (b) Partitioning of the interaction rate between a QE and the quantized SPCM states into the coherent, $\lambda$, and dissipative, $\eta$, quantum exchange rates.}
\label{Fig:PC}
\end{figure}

In this paper, we explore possibilities of engineering quantum critical properties of plasmonic cavities by examining the non-equilibrium phase diagram and associated photon emission spectra with respect to the nature and strength of the surface plasmon - QE interactions. To do so, we generalize a driven-dissipative Dicke model describing an ensemble of identical two-level QEs coupled to a single bosonic surface plasmon mode referred to below as the surface plasmon cavity mode (SPCM) [Fig.~\ref{Fig:PC}(a)]. We incorporate into the model two distinct SPCM-QE interaction pathways which we illustrate in Fig.~\ref{Fig:PC}(b): a  coherent coupling, $\lambda$, which stems from a  direct, e.g., dipole-dipole interaction between the SPCM and a QE and a dissipative coupling, $\eta$,  which is facilitated by coupling to a reservoir,  which may be either a photon continuum or a dark surface plasmon mode (illustrated in Fig.~\ref{Fig:PC}(a)) interacting in the near-field with both the SPCM and QEs. {\color{black} The energy is injected into the cavity incoherently, e.g., via electrical pumping or optical excitation of the QE high excited states followed by phonon assisted relaxation to the ``active" excited state.}  The SPCM is coupled to the photon continuum facilitating the cavity emission.

The dissipative coupling via photon reservoir was introduced by Lehmberg as the off-diagonal radiative decay terms in the Lindblad operator to describe the superradiant emission from an ensemble of two-level atoms~\cite{LehmbergPRA:1970}. Subsequently, this approach has been widely used to study the superradiant emission in a large variety of systems~\cite{SpanoJCP:1989,Pustovit:10,ShammahPRA:2018}. Coupling via a reservoir has been introduced in cascaded quantum systems~\cite{PlenioRMP:1998,MetelmannPRX:2015} and shown to be important for quantum information applications~\cite{VerstraeteNatPhys:2009}. The dissipative coupling is also shown to facilitate quantum entanglement~\cite{KastoryanoPRL:2011,ReiterPRL:2016,HaoLiJCP:2019} and can be employed to steer open quantum system into non-trivial states~\cite{DiehlNatPhys:2008,HamaPRL:2018}.   

The Dicke model provides a fundamental model of cavity QED and has been studied in a context of superradiant phase transition both in equilibrium and non-equilibrium regimes.\cite{Kirton_AdvQT:2019} Its generalized version with imbalanced rotating and counter-rotating terms has revelled a rich phase diagram allowing for the superradiant and various lasing states~\cite{Kirton_NJP:2018,Shchadilova_arXiv:2018,Kirton_AdvQT:2019}. In this paper, we demonstrate that the interplay of the coherent and dissipative interactions incorporated into the plasmonic Dicke model results in the extension of the superradiant and regular lasing phases to the dissipative coupling region of the phase diagram and shows an emergent lasing state without population inversion (referred below as lasing without inversion) having phase boundaries with the superradiant and normal steady states.  We demonstrate that the associated photon emission spectra carry distinct spectroscopic signatures of the various phases.  

{\color{black} The paper organized as follows.  Generalization of the plasmonic Dicke model accounting for the dissipative coupling is presented in Sec.~\ref{Sec:SPDicke}. In Sec.~\ref{Sec:NeqPhD} the mean field and second moment equations of motion are employed to calculate the non-equilibrium phase diagram and to identify the steady states. Associated photon emission spectra are analyzed in Sec.~\ref{Sec:PEms} and the conclusions are drown in Sec.~\ref{Sec:Conc}.}

\section{Generalized Dicke Model} 
\label{Sec:SPDicke}

The SPCM cavity mode (Fig.~\ref{Fig:PC}) is described by Bose operators, $\{\psi^\dag,\psi\}$. Each two-level QE occupying site $n=\overline{1,{\cal N}_o}$ is characterized by a set of spin operators $\{\hat s_n^\pm=\hat s_n^x \pm i\hat s^y_n$, $s_n^z\}$, related to the Pauli SU(2) operators as $\hat s_n^{j}=\frac{1}{2}\hat\sigma_n^{j}$ with $j=x,y,z$. 
Assuming that the QEs and the SPCM have the same resonance energy, $\omega_o$, and the same coherent quantum exchange rate, $\lambda$, we describe the system by the following Dicke Hamiltonian given in units of $\hbar$ 
\begin{eqnarray}
\label{Hpl-Dicke}
\hat H_D &=& \omega_o\hat\psi^\dag\hat\psi 
		+\omega_o\left(\sum\limits_{n=1}^{{\cal N}_o}\hat s_n^z+\frac{{\cal N}_o}{2}\right) 
\\\nonumber &~&
		+\lambda\left(\hat\psi+\hat\psi^\dag\right)\sum\limits_{n=1}^{{\cal N}_o}\left(\hat s_n^{-}+\hat s_n^{+}\right).
\end{eqnarray}

By taking into account that the plasmonic cavity (Fig.~\ref{Fig:PC}) is an open quantum system, we introduce a density operator $\hat\rho$ projected on the SPCM and QE space whose time evolution is described by a Liouville equation 
\begin{eqnarray}
\label{Liov-eq-Dicke}
\partial_t\hat\rho &=& -i\left[\hat H_D,\hat\rho\right] 
    + \Gamma_{sp}\hat{\cal D}_{\hat\psi}[\hat\rho]
    + \frac{\gamma_\uparrow}{2}\sum\limits_{n=1}^{{\cal N}_o}\hat{\cal D}_{\hat s_n^{+}}[\hat\rho]
\\\nonumber&~&
	+ \frac{\gamma_\downarrow}{2}\sum\limits_{n=1}^{{\cal N}_o}\hat{\cal D}_{\hat s_n^{-}}[\hat\rho]
	+ \gamma_\phi\sum\limits_{n=1}^{{\cal N}_o}\hat{\cal D}_{\hat s_n^z}[\hat\rho] 
	+\eta\sum\limits_{n=1}^{{\cal N}_o}\hat{\cal D}_{\hat\psi,\hat s_n^{+}}[\hat\rho].
\end{eqnarray}
Here, the Lindblad superoperator $\hat{\cal D}_{\hat{\cal O}}[\hat\rho] = (2\hat{\cal O}\hat\rho\hat{\cal O}^\dag-\hat{\cal O}^\dag\hat{\cal O}\hat\rho-\hat\rho\hat{\cal O}^\dag\hat{\cal O})$ with $\hat{\cal O}=\{\hat\psi, \hat s^{-}_n, s^{+}_n,\hat s_n^z\}$ acts either within the SPCM or QE subspaces. It describes the SPCM population decay with the rate $2\Gamma_{sp}$, QEs' population decay with the rate $\gamma_\downarrow$, population gain due to an incoherent pump with the rate $\gamma_\uparrow$, and a pure dephasing with the rate $\gamma_\phi$. All introduced population decay rates include both the radiative and non-radiative contributions. The last term in Eq.~(\ref{Liov-eq-Dicke}) partitions the {\em dissipative} SPCM and QE interaction with the rate $\eta$ using the Lindblad operator~\footnote{In general, a Lindblad superoperator, $\gamma\sum_n \hat{\cal D}_{\alpha\hat\psi+\beta\hat s^{-}_n}[\hat\rho]$, for the interacting SPCM and QE contains three independent coupling parameters:  $\alpha$, $\beta$, $\gamma$. Here, we use transformed {\em independent} parameters: $\Gamma_{sp}\sim \alpha^2\gamma$, $\gamma_{\downarrow}\sim \beta^2\gamma$, $\eta = \alpha\beta\gamma$.}
\begin{eqnarray}
\label{LR-SP-QE}
\hat{\cal D}_{\hat\psi,\hat s_n^{+}}[\hat\rho]&=& 2\hat\psi\hat\rho\hat s_n^{+}
			-\hat s_n^{+}\hat\psi\hat\rho-\hat\rho\hat s_n^{+}\hat\psi
\\\nonumber&+&
2\hat s_n^{-}\hat\rho\hat\psi^\dag
			-\hat\psi^\dag\hat s_n^{-}\hat\rho-\hat\rho\hat\psi^\dag\hat s_n^{-}. 
\end{eqnarray}
Eqs.~(\ref{Hpl-Dicke})--(\ref{LR-SP-QE}) constitute our generalization of a driven-dissipative plasmonic Dicke model. 

{\color{black} Microscopic derivation of this model with the SPCM-QE interactions facilitated by the photon continuum is provided in Appendix~\ref{Appx:PhBath}. Associated expressions for the coherent and incoherent coupling rates are given by Eqs.~\eqref{lmbd_spqe} and \eqref{eta_spqe} complemented by Eqs.~\eqref{lambda_ab} and \eqref{eta_ab}, respectively. These equations naturally account for the photon retardation effects allowing one to treat plasmonic cavities with linear size of the order of an optical wavelength, the regime when  conventional near-field dipole-dipole approximation breaks down. In the near-field limit, the dissipative interaction (Eq.~\eqref{eta_ab}) becomes week, i.e., scales with the distance $r$ between the dipole as $r^{-1}$. Therefore, dropping the dissipative term in the near-field limit, we recover a plasmonic cavity model used in Ref.~\cite{ZasterJPCS:2016} to study equilibrium exciton-plasmon polariton condensation. As demonstrated in Appendix~\ref{Appx:2SPM}, an alternative way to achieve dissipative interaction, including the near-field limit, is to facilitate the SPCM-QEs interactions via a broad dark plasmon mode.}

\section{Nonequilibrium Phase Diagram}
\label{Sec:NeqPhD}

\subsection{Mean-field analysis} 
\label{Sec:MFA}

The following mean-field equations of motion directly follow from Eqs.~(\ref{Hpl-Dicke})--(\ref{LR-SP-QE}) 
\begin{eqnarray}
\label{mf-SPcoh}
\partial_\tau {\psi} &=& -\left(i+\bar\Gamma_{sp}\right){\psi}
 -2i{\cal N}_o\bar\lambda\text{Re}[s_{-}]-{\cal N}_o\bar\lambda\bar\eta s_{-}, 
\\\label{mf-scoh}
\partial_\tau s_{-}&=&-\left(i+\bar\Gamma_o\right)s_{-}
	+4i\bar\lambda s_z \text{Re}\left[\psi\right]+2\bar\lambda\bar\eta s_z\psi,		
\\\label{mf-spop}
\partial_\tau s_z &=& - \bar\gamma_o\left(s_z-\frac{d_o}{2}\right)
	-4\bar\lambda\text{Im}\left[s_{-}\right]\text{Re}\left[\psi\right]
\\\nonumber &~&	
	-2\bar\lambda\bar\eta\text{Re}\left[\psi^*s_{-}\right],		
\end{eqnarray}
where the scalar variables are $\psi=\langle\hat\psi\rangle$, $s_{-}=\langle\hat s_{-}\rangle$, and $s_z=\langle\hat s_z\rangle$ with the normalized per site spin operators, $\hat s_{\pm,z}=\sum_{n}(\hat s_n^{\pm,z})/{\cal N}_o$, and the brackets defined as $\langle\hat{\cal O}\rangle=\text{tr}(\hat{\cal O}\hat\rho)$. $\tau=\omega_o t$ is a dimensionless time variable. $\bar\lambda = \lambda/\omega_o$  [$\bar\eta=\eta/\lambda$] is normalized coherent [dissipative] coupling rate. The normalized SPCM dephasing rate is $\bar\Gamma_{sp}=\Gamma_{sp}/\omega_o$ and the QE dephasing [total population decay] rate is $\bar\Gamma_o=(\gamma_\downarrow/2+\gamma_\uparrow/2+\gamma_\phi)/\omega_o$  [$\bar\gamma_o=(\gamma_\downarrow+\gamma_\uparrow)/\omega_o$]. Finally, $d_o=(\gamma_\uparrow-\gamma_\downarrow)/(\gamma_\uparrow+\gamma_\downarrow)$ is the population inversion parameter.

\begin{figure*}[t]
\includegraphics[width=0.8\textwidth]{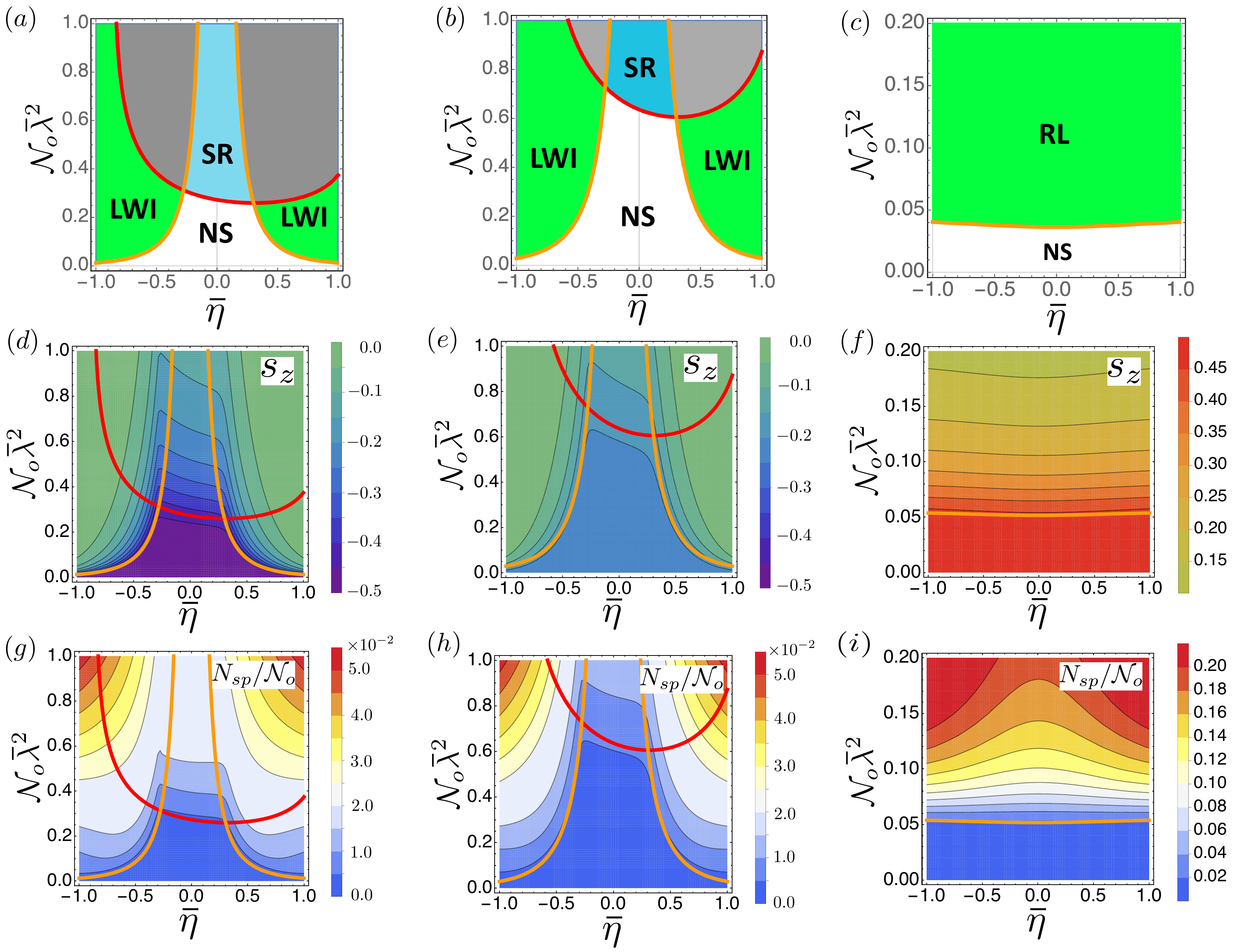}
\caption{(a)-(c): Mean-field phase diagrams marking the normal (NS), superradiant (SR), lasing without inversion (LWI) and regular lasing (RL) states. (d)-(f) The steady state projection on the QE population inversion $s_z$ and (g)-(i) reduced SPCM population $N_{sp}/{\cal N}_o$  calculated for ${\cal N}_o=200$ using Eqs.~\eqref{eqm-Nsp-nrzd}--\eqref{eqm-C+-nrzd}. In the left, central, and right columns  $d_o=-1, -0.4, 1$, respectively. Red [orange] curves designate ${\cal N}_o\bar\lambda_s^2(\eta)$ [${\cal N}_o\bar\lambda_l^2(\eta)$] boundaries. In all calculations $\bar\Gamma_{sp}=0.3$, $\gamma_\downarrow/\omega_o=0.002$, and $\gamma_\phi/\omega_o=0.02$.}
\label{Fig:PHD}
\end{figure*}

The normal state is the trivial steady state of Eqs.~(\ref{mf-SPcoh})--(\ref{mf-spop}) characterized by the QE population inversion $s^\text{ns}_z=d_o/2$ and the absence of the SPCM and QE coherences, $s_{-}^\text{ns}=\psi^\text{ns}=0$. 
The fluctuations of the normal state coherences,  $\delta\psi=\psi-\psi^\text{ns}$ and $\delta s_{-}= s_{-}-s_{-}^\text{ns}$, satisfy linearized Eqs.~(\ref{mf-SPcoh}) and (\ref{mf-scoh}) represented in the matrix form
\begin{eqnarray}
\label{EqM-def}
\partial_{\tau} \textbf{\em v} = {\cal M} \textbf{\em v},		
\end{eqnarray}
with the vector $\textbf{\em v}=[\delta\psi, \delta\psi^*, \delta s_{-}, \delta s_{-}^*]^\texttt{T}$ and the stability matrix~\footnote{The linearized Eq.~\eqref{mf-spop} for the QE inversion fluctuations, $s_z=s_z-\delta s_z$, is uncoupled from Eqs.~\eqref{EqM-def} and \eqref{Ms-def} and has no effect on the normal state stability.},
\begin{eqnarray}
\label{Ms-def}
&~&{\cal M}=
\\\nonumber &~&
\begin{bmatrix} 
 -i-\bar\Gamma_{sp} & 0 &-{\cal N}_o\bar\lambda(i+\bar\eta)	& -i{\cal N}_o\bar\lambda\\
0	& i-\bar\Gamma_{sp} &i{\cal N}_o\bar\lambda	&{\cal N}_o\bar\lambda(i-\bar\eta)\\
2s_z\bar\lambda(i+\bar\eta)& 2is_z\bar\lambda&  -i-\bar\Gamma_o &0 \\
-2is_z\bar\lambda& -2s_z\bar\lambda(i-\bar\eta) & 0&i-\bar\Gamma_o
	\end{bmatrix}.		
\end{eqnarray}
To identify a phase diagram of our model as the function of dimensionless coherent, ${\cal N}_o\bar\lambda^2$, and dissipative, $\bar\eta$, coupling parameters, we look for the normal state instabilities of Eq.~\eqref{EqM-def}. Specifically, we calculate the eigenvalues of the matrix ${\cal M}$ with $s_z=s_z^\text{ns}$ and check if at least one of them acquires positive real part.

Figure~\ref{Fig:PHD} presents a phase diagram of the calculated steady states at {\em no pumping} $d_o=-1$ (panel~(a)), and {\em below} the population inversion with $d_o=-0.4$ (panel~(b)) where three distinct instability regions are identified. The cyan region
denotes a phase associated with a single positive eigenvalue of the stability matrix that passes through zero at the boundary with the normal state (NS). This is a signature of the pitchfork bifurcation characterizing a superradiant phase transition~\cite{Kirton_AdvQT:2019} breaking $\mathbb{Z}_2$ symmetry, $\psi \rightarrow -\psi$ and $s_- \rightarrow -s_-$, of Eqs.~(\ref{mf-SPcoh})--(\ref{mf-spop}). Accordingly, the critical coherent coupling for this transition can be evaluated as the root of $\det[{\cal M}]=0$ resulting in  
\begin{eqnarray}
\label{lmbc-superrad}
{\cal N}_o\bar\lambda_s^2(\eta)&=&\frac{P(\bar\eta)}
{\bar\eta^4d_o}
\left[\sqrt{1-\frac{\bar\eta^4(1+\bar\Gamma_{sp}^2)(1+\bar\Gamma_o^2)}{P^2(\bar\eta)}}-1\right],
\;\;
\end{eqnarray}
with $P(\bar\eta)=2+2\bar\eta(\bar\Gamma_o+\bar\Gamma_{sp})+\bar\eta^2(\bar\Gamma_o\bar\Gamma_{sp}-1)$. For vanishing dissipative coupling $\bar\eta\rightarrow 0$, Eq.~\eqref{lmbc-superrad} recovers a critical coupling ${\cal N}_o\bar\lambda_c^2=(1+\bar\Gamma^2_{sp})(1+\bar\Gamma^2_o)/(-8 s_z^\text{ns})$ for the superradiant phase transition in the open Dicke model.\cite{Kirton_AdvQT:2019} The red curves in Fig.~\ref{Fig:PHD}~(a) and (b) due to Eq.~\eqref{lmbc-superrad}  marks well the boundaries between the normal state and the identified superradiant (SR) region.

The green regions in Fig.~\ref{Fig:PHD}~(a) and (b) mark the instabilities characterized by simultaneous appearance of the real positive part for two complex eigenvalues of ${\cal M}$. This is a signature of a critical Hopf bifurcation. In the Tavis-Cummings model, i.e., the rotating wave limit of the Dicke model, such an instability points to the lasing phase transition,
which  breaks $U(1)$ gauge symmetry $\psi\rightarrow e^{i\bar\omega_l\tau}\psi $, $s_-\rightarrow e^{i\bar\omega_l\tau} s_-$ where $\bar\omega_l=\omega_l/\omega_o$ is a dimensionless lasing frequency.\cite{Kirton_AdvQT:2019} By applying the rotating wave approximation to Eq.~\eqref{mf-SPcoh}--\eqref{mf-spop}, we recover a generalized Tavis-Cummings model (see Appendix~\ref{Appx:GTC}) with the critical parameter 
\begin{eqnarray}
\label{lmbc-laser}
{\cal N}_o\bar\lambda_l^2(\eta)&=&
\frac{(\bar\eta^2-1)\left(\bar\Gamma_o+\bar\Gamma_{sp}\right)^2}{8\bar\eta^2d_o}
\\\nonumber&\times&
\left[1\pm\sqrt{1+\frac{16\bar\eta^2\bar\Gamma_{sp}\bar\Gamma_o}
{(\bar\eta^2-1)^2\left(\bar\Gamma_o+\bar\Gamma_{sp}\right)^2}}~\right].
\end{eqnarray}

Such a critical parameter exists {\em below} the population inversion, $-1\leq d_o<0$, and the {\em non-vanishing} dissipative coupling if the plus (minus) sign set in front of the square root for $0<\bar\eta^2<1$ ($\bar\eta^2 > 1$). The orange curves in Fig.~\ref{Fig:PHD}~(a) and (b) due to Eq.~\eqref{lmbc-laser} mark well the boundaries between the normal state and the green region. Accordingly, we identify the latter as a lasing states without population inversion facilitated by the dissipative SPCM-QE interaction. The crossover region (gray in Fig.~\ref{Fig:PHD}~(a) and (b)) exhibits an interplay of both superradiant and lasing instabilities resulting in properties to be clarified below.       

Above the population inversion, $0<d_o\leq 1$, our generalized Dicke model recovers the regular lasing state as marked in Fig.~\ref{Fig:PHD}~(c) by green. Variation of the critical boundary there  is in good agreement with the predictions (orange line) of Eq.~\eqref{lmbc-laser} where the plus (minus) sign is adopted for $\bar\eta^2>1$ ($\bar\eta^2<1$). In this case, a limit of $\bar\eta\rightarrow 0$ exists resulting in a well-known form of the QE-cavity coupling ${\cal N}_o\bar\lambda_l^2=\bar\Gamma_o\bar\Gamma_{sp}/d_o$ at the  lasing threshold.\cite{Gardiner_Qnoise:2004}

\subsection{Second moment analysis} 
\label{Sec:SMA}

To go beyond the mean-field analysis of the instability regions shown in Fig.~\ref{Fig:PHD}~(a)-(c), we employ a set of equations of motion truncated at the level of all independent operator pairs. These averages represent a complete set of second moments of the density operator in the space of  interacting QE-SPCM states as introduced in Sec.~\ref{Sec:SPDicke}. In the literature such an approximation is also known as the maximum entropy two-particle factorization scheme~\cite{Mukamel_book:1995} or the second cumulant approximation~\cite{Kirton_NJP:2018,Kirton_AdvQT:2019}.     

{\color{black} Starting with Eqs.~(\ref{Hpl-Dicke})--(\ref{LR-SP-QE}), we derive the following equations of motion for the SPCM population, $N_{sp}=\langle\hat\psi^\dag\hat\psi\rangle$, and double coherence, $C_{sp}=\langle\hat\psi\hat\psi\rangle$, 
\begin{eqnarray}
\label{eqm-Nsp-nrzd}
\partial_\tau N_{sp} &=& 
	-2\bar\Gamma_{sp}N_{sp}
	-2{\cal N}_o\bar\lambda\text{Im}\left[c_{-sp}+c_{+sp}\right]
\\\nonumber&~&	
	-2{\cal N}_o\bar\lambda\bar\eta\text{Re}\left[c_{+sp}\right],
\\\label{eqm-Csp-dmls}
\partial_\tau C_{sp}&=&
	-2\left(i+\bar\Gamma_{sp}\right)C_{sp}
	-2i{\cal N}_o\bar\lambda\left[c_{-sp}+c_{+sp}\right]
\\\nonumber&~&	
	-2{\cal N}_o\bar\lambda\bar\eta c_{-sp}.
\end{eqnarray}
To close the set, equations of motion for the SPCM-QE coherences, $c_{\pm sp}=\sum_n\langle \hat s^\pm\hat\psi\rangle/{\cal N}_o$,  
\begin{eqnarray}
\label{eqm-C-sp-nrzd}
\partial_\tau c_{- sp} &=& -\left(2i+\bar\Gamma_{sp}+\bar\Gamma_o\right)c_{- sp}
\\\nonumber &~&		
		-i\bar\lambda\left[({\cal N}_o-1)(c_{+-}+c_{--})
\right.\\\nonumber &~&\left.
		-2s_z\left(C_{sp}+N_{sp}\right)+1/2-s_z\right]
\\\nonumber &~&		
	-\bar\lambda\bar\eta\left[ ({\cal N}_o-1)c_{--}-2s_z C_{sp}\right],
\\\label{eqm-C+sp-nrzd}
\partial_\tau c_{+ sp} &=& -\left(\bar\Gamma_{sp}+\bar\Gamma_o\right)c_{+ sp}
\\\nonumber &~&	
	-i\bar\lambda\left[\left({\cal N}_o-1\right)\left(c_{+-}+c_{--}^*\right)
\right.\\\nonumber &~&\left.
	+2s_z\left(C_{sp}+N_{sp}\right)	+1/2+s_z\right]
\\\nonumber &~&	
	-\bar\lambda\bar\eta\left[\left({\cal N}_o-1\right)c_{+-}-2s_z N_{sp}+1/2+s_z\right],	
\end{eqnarray}
average QE population $s_z$, QE double coherence $c_{--}=\sum_{n\neq n'}\langle \hat s_n^{-} \hat s_{n'}^{-}\rangle/({\cal N}_o({\cal N}_o-1))$, and the intersite coherence $c_{+-}=\sum_{n\neq n'}\langle \hat s_n^{+} \hat s_{n'}^{-}\rangle_{n\neq n'}/({\cal N}_o({\cal N}_o-1))$, 
\begin{eqnarray}
\label{eqm-Jpop-nrzd}
\partial_\tau s_z &=& -\bar\gamma_o\left(s_z-d_o/2\right)
		-2\bar\lambda\text{Im}\left[c_{-,sp}-c_{+,sp}\right]
\\\nonumber &~&		
		-2\bar\lambda\bar\eta \text{Re}\left[c_{+sp}\right],
\\\label{eqm-C--nrzd}
\partial_\tau c_{--} &=& -2\left(i+\bar\Gamma_o\right)c_{--} 
	+4i\bar\lambda s_z\left[c_{- sp}+c_{+ sp}^*\right]
\\\nonumber &~&	
						+4\bar\lambda\bar\eta s_z c_{-sp},
\\\label{eqm-C+-nrzd}
\partial_\tau  c_{+-} &=& -2\bar\Gamma_o c_{+-} 
			-4\bar\lambda s_z\text{Im}[c_{+ sp}-c_{- sp}]
\\\nonumber &~&			
				+4\bar\lambda\bar\eta s_z \text{Re}\left[c_{+sp}\right],							
\end{eqnarray}
are introduced, respectively. $C_{sp}$, $c_{\pm sp}$, and $c_{- -}$ are complex quantities and the associated equations should be complimented by their complex conjugates.\footnote{In the limit of vanishing dissipative coupling, $\bar\eta=0$, Eqs.~\eqref{eqm-Nsp-nrzd}--\eqref{eqm-C+-nrzd} recover the results of Ref.~[\onlinecite{Kirton_NJP:2018}].}

In the absence of the incoherent pumping, $d_o=-1$, Eqs.~\eqref{eqm-Nsp-nrzd}--\eqref{eqm-C+-nrzd} describe the relaxation dynamics of the elementary excitations (introduced as initial conditions) towards the ground state whose projection on the QE population inversion, $s_z$, and the SPCM population, $N_{sp}=\langle\hat\psi^\dag\hat\psi\rangle$, are shown in panels~(d) and (g) of Fig.~\ref{Fig:PHD}, respectively. In agreement with the mean-field calculations, the normal state is characterized by the QE ground state, $s^\text{ns}_z=-1/2$, and empty plasmonic cavity, $N_{sp}=0$.    However, panels~(d) and (g) of Fig.~\ref{Fig:PHD} show that presence of the counter-rotating term, $\hat H_{CR}=\lambda\sum_{n}(\hat\psi\hat s_n^{-}+\hat\psi^\dag\hat s_n^{+})$, in the Hamiltonian~\eqref{Hpl-Dicke} results in QE and SPCM excitations as the coupling parameters cross the critical boundaries. We also observed that above the critical boundaries, the coherences entering Eqs.~\eqref{eqm-Nsp-nrzd}--\eqref{eqm-C+-nrzd} become non-zero (not shown in the plot). This is an indication that the ground state associated with the symmetry breaking is a correlated state of the QEs and SPCM. Since the basis of the bare QE and SPCM states is no longer the eigenbasis, projections of the ground state density matrix on the QE and SPCE populations show number of {\em virtual} excitations present in the ground state. Accordingly, Eqs.~\eqref{eqm-Nsp-nrzd}--\eqref{eqm-C+-nrzd} describe relaxation of the elementary excitations to such a correlated ground state avoiding a problem of unphysical plasmon generation in the ground state $\hat H_{CR}$ discussed in Ref.~\cite{KockumNatRevPhys:2019}. }

When the energy is incoherently supplied to the QEs, panels~(e), (f) and (h), (i) of Fig.~\ref{Fig:PHD} show that still in agreement with the mean-field theory the normal state regions are characterized by $s^\text{ns}_z=d_o/2$ and $N_{sp}=0$, respectively. Observed variation of $s_z$ and a build up of $N_{sp}$ within associated instability regions (panels~(b), (c)) indicate the non-equilibrium phase transitions to the correlated QE-SPCE states. Interestingly, the crossover region (gray in panels~(a), (b)) shows a clear phase boundary with the superradiant state in the associated QE and SPCM population plots. However, no signature of the phase boundary with the lasing state without inversion is seen suggesting that lasing features might be expected in this region. 

\section{Photon Emission Properties} 
\label{Sec:PEms}

Having established the phase diagram, we further examine photon emission spectra of the identified steady states. {\color{black} Keeping in mind that the SPCM emission rate exceeds the same quantity for QEs by orders of magnitude, in Appendix~\ref{Appx:InOut}, we use the input-output formalism to connect the photon energy emission spectrum, $S(\delta\bar\omega)$, measured by a photo-detector with the Fourier transformed SPCM operator auto-correlation function.} Specifically,   
\begin{eqnarray}
\label{Sw}
S(\delta\bar\omega)=(1+\delta\bar\omega)^4\int_{-\infty}^\infty  \frac{d\tau}{2\pi}  \langle \hat\psi^\dag(\tau)\hat\psi(0)\rangle e^{i\delta\bar\omega\tau},
\end{eqnarray}
with the dimensionless frequency detuning $\delta\bar\omega=(\omega-\omega_o)/\omega_o$, and the prefactor $(1+\delta\bar\omega)^4$ reflecting scaling of the photon density of states and the emitted photon energy. {\color{black} Noteworthy that the spectral distribution of the number of emitted photons can be obtained by normalizing $S(\delta\bar\omega)$ per photon energy which changes the frequency prefactor to $(1+\delta\bar\omega)^3$. The cubic scaling of the prefactor reflects the scaling of the spontaneous decay rate determined by the vacuum photon density of states.}

 \begin{figure}[t]
\includegraphics[width=0.4\textwidth]{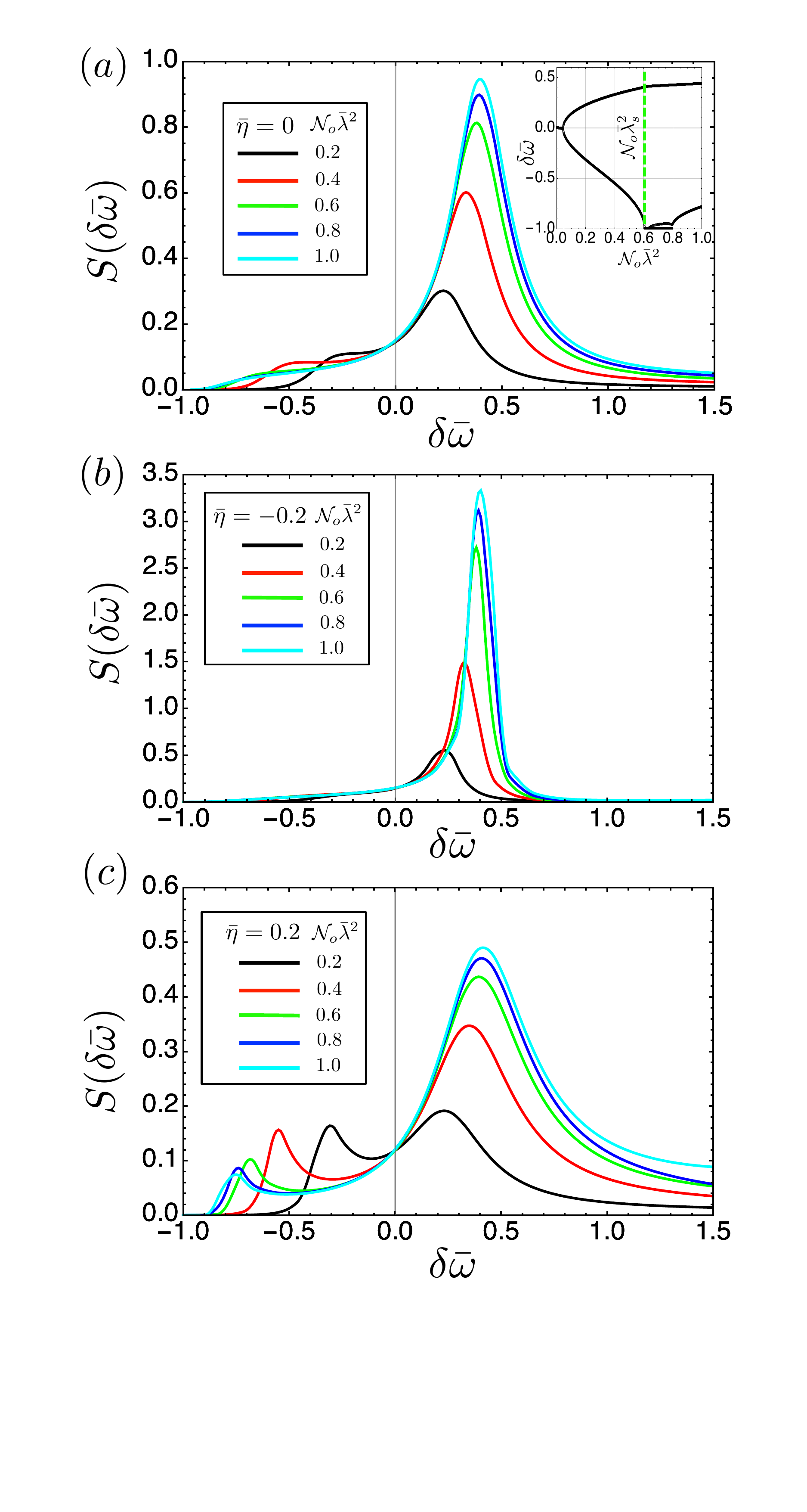}
 \caption{\color{black} The photon energy emission spectra below population inversion, $d_o=-0.4$. The coherent coupling, changes from the normal ${\cal N}_o\bar\lambda^2<0.6$ to the superradiant ${\cal N}_o\bar\lambda^2 \geq 0.6$ region and the incoherent coupling is set to (a) $\bar\eta=0$, (b) $\bar\eta=-0.2$, and (c) $\bar\eta=0.2$. The inset: mean-field polariton spectrum for $\bar\eta=0$.}
 \label{Fig:EPS}
 \end{figure}

Applying the quantum regression theorem~\cite{Gardiner_Qnoise:2004}, the correlation function in Eq.~\eqref{Sw} is calculated perturbatively via numerical solution of Eqs.~\eqref{EqM-def} and \eqref{Ms-def} with  $\textbf{\em v}(\tau)=[\langle\hat\psi(\tau)\hat\psi(0)\rangle, \langle\hat\psi^\dag(\tau)\hat\psi(0)\rangle, \langle \hat s_{-}(\tau)\hat\psi(0)\rangle, \langle \hat s_{+}(\tau)\hat\psi(0)\rangle]^\texttt{T}$. The second moment steady states are used as the initial conditions for $v(0)$ and the steady state $s_z$ is used to parameterize ${\cal M}$. {\color{black} According to the adopted formalism the photon emission occurs as a result of relaxation of the SPCM-QM correlated elementary excitations to the steady/ground state accounting for virtual SPCM and QE excitations above the critical boundaries. This allows us to eliminate unphysical photon emission by such virtual states discussed in Ref.~\cite{KockumNatRevPhys:2019}.}

Figure~\ref{Fig:EPS}(a) compares the emission spectra in transition from the normal to the superradiant steady state at no dissipative coupling, i.e., the $\bar\eta=0$ slice of the phase diagram in Fig.~\ref{Fig:PHD}~(b). Each curve has two features: one characterized by a positive and the other by a negative detuning. The energy splitting between the features is in quantitative agreement with the spectrum of the elementary excitations~\cite{HaoLi-eph:2018}, namely the surface plasmon-exciton polaritons, shown in the inset. 

As the coherent coupling passes through the superradiant critical value (green dash in the inset), the lower polariton branch passes through a gap at zero photon energy ($\delta\bar\omega=-1$)~\cite{KopylovPRA:2013}. Emission within this spectral branch is highly suppressed due to the decrease in the photon energy and the density of states (the prefactor in Eq.~\eqref{Sw}). Therefore, a subtle lower polariton behavior near the critical point is not resolved in the spectra. As a results, the emission at the superradiant phase transition occurs from the upper polariton branch. 

{\color{black} Small non-vanishing values of the incoherent coupling, $\bar\eta$, {\em below} the threshold for the lasing without inversion, result in the emission spectrum modifications shown in panels (b) and (c) of Fig.~\ref{Fig:EPS} Specifically, negative values of $\bar\eta$ (panel (b)) result in the emission line narrowing and strong suppression of the lower polariton peak. In contrast, positive values of $\bar\eta$ (panel (c)) increase the line broadening and enhance the lower polariton features.}

\begin{figure}[t]
\includegraphics[width=0.4\textwidth]{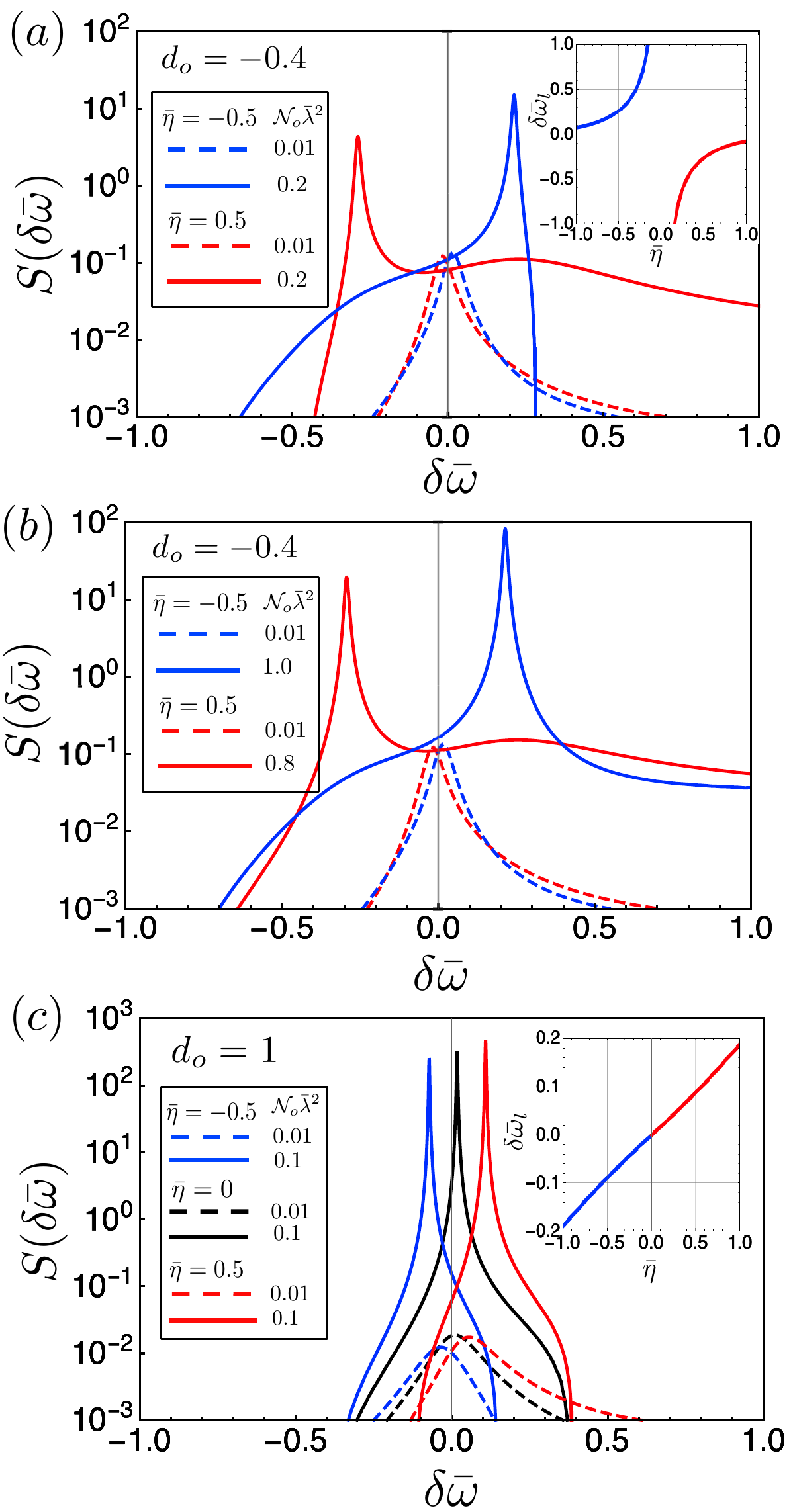}
 \caption{\color{black} The photon emission spectra: (a) and (b) Solid (dashed) lines mark lasing state without inversion (normal state) emission with the coupling threshold ${\cal N}_o\bar\lambda_l = 0.2$. The inset to panel (a) shows mean-field calculated lasing frequency shift. (c) Same as (a) and (b) but for the regular lasing above the inversion characterized by the coupling threshold ${\cal N}_o\bar\lambda_l = 0.04$.}
 \label{Fig:ELS}
 \end{figure}

Figure~\ref{Fig:ELS}(a) compares emission spectra for $\bar\eta=\pm 0.5$ values of the dissipative coupling below (dashed line) and within (solid line) the lasing without inversion phase shown in Fig.~\ref{Fig:PHD}~(b). A sharp characteristic features appear above the lasing critical coupling. To identify these features, we have calculated a normalized lasing frequency shift $\delta\bar\omega_l=(\omega_l-\omega_o)/\omega_o$ using the generalized Tavis-Cummings model (see Appendix~\ref{Appx:GTC}) 
\begin{eqnarray}
\label{dwl-eta}
\delta\bar\omega_l(\eta)=-d_o\frac{2\bar\eta {\cal N}_o\bar\lambda_l^2(\eta)}{\bar\Gamma_o+\bar\Gamma_{sp}},
\end{eqnarray}
where ${\cal N}_o\bar\lambda_l^2(\eta)$ given by Eq.~\eqref{lmbc-laser}. According to this expression, the peak shift is determined by the dissipative coupling parameter $\bar\eta$. 

The lasing peak shift evaluated according to Eq.~\eqref{dwl-eta} is plotted in the inset to Fig.~\ref{Fig:ELS}(a). Comparison shows that the spectral positions of the sharp emission peaks are in quantitative agreement with the predictions of the Tavis-Cummings model allowing us to identify them as the lasing peaks. Spectra calculated within the crossover region (panel~(b)) demonstrate the same trends as the spectra in panel (a). This confirms our assumption that the crossover region could demonstrate the lasing features. Finally, Fig.~\ref{Fig:ELS}(c) compares the emission spectra for the regular lasing regime identified in Fig.~\ref{Fig:PHD}~(c). Compared to the broad spectral distribution of the photons (dashed lines) emitted below the lasing critical coupling, the spectra associated with the lasing phase show sharp monochromatic emission feature. The spectral position of the latter features are in quantitative agreement with the predictions of Eq.~\eqref{dwl-eta} plotted in the inset.   

Comparing the insets in panels (a) and (c) of Fig.~\ref{Fig:ELS}, we notice that given the same sign of the dissipative coupling strength, the lasing frequency shift for the lasing without inversion and regular lasing regimes shows opposite signs. This trend resembles the behavior predicted for the counter-rotating (also known as the inverted) and regular lasing regimes using a generalization of the Dicke model reported in Refs.~\cite{ZhiqiangOtp:2017,Kirton_NJP:2018,Shchadilova_arXiv:2018,Kirton_AdvQT:2019}. Furthermore, the counter-rotating lasing occurs below the population inversion, $d_o<0$, and has a phase boundary with the superradiant state. 

However, the physical mechanisms leading to the lasing without inversion reported here and the counter-rotating lasing are different. In the former case, a gain occurs as the result of the reservoir facilitated energy flow from the SPCM to the QEs. This results in the energy supply to the QEs that overrides their losses. In the case of the counter-rotating lasing, a cavity coherent emission is facilitated by the counter rotating,  $\hat H_{CR}=\lambda\sum_{n}(\hat\psi\hat s_n^{-}+\hat\psi^\dag\hat s_n^{+})$, term. To reach the counter-rotating lasing an imbalance between the rotating and the counter rotating terms is required. As demonstrated for atoms in a high finesse optical cavity, such an imbalance can be engineered using coherent Raman processes due to external laser fields.\cite{DimerPRA:2007,ZhiqiangOtp:2017} In contrast, the lasing without inversion predicted in this paper does not originate from the imbalanced counter-rotating terms but rather emerges due to the SPCM-QM interaction facilitated by a reservoir.     

\section{Conclusions}
\label{Sec:Conc}
 
Using the mean-field and second moment analysis we identified the superradiant, lasing without inversion and regular lasing states appearing in the phase diagram of driven-dissipative plasmonic Dicke model as a result of interplay between the coherent and dissipative SPCM-QE interaction pathways. In the limit of no dissipative coupling, we recover the results for the open Dicke and Tavis-Cummings model reported in the literature. Inclusion of the dissipative coupling extends the superradiant and regular lasing states into the dissipative coupling region and results in the emergent lasing without inversion state. The latter stems from the incoherent energy transfer between the SPCM and the QEs. The calculated emission spectra show quantitative agreement with the predictions of the the mean-field theory. The trends for the peak variations and the line shape behavior with respect to the strengths of the dissipative coupling can be used for a spectroscopic identification of the predicted states.

\begin{acknowledgments}

The work at Los Alamos National Laboratory is supported by the LANL LDRD program. AP also acknowledges support provided by the LANL IMS to visit Durham University and University of St. Andrews where part of this work was done.   OR acknowledges startup funds provided by Fordham University.  
The work at the University of Houston was funded in part by the  National Science Foundation 
(CHE-1664971, 
CHE-1836080), 
 and the Robert A. Welch Foundation (E-1337).
ERB also acknowledges the Leverhulme Trust for 
support at Durham University where part of this work was completed.
We thank Jonathan Keeling and Peter W. Milonni for stimulating discussions.  
\end{acknowledgments}

\appendix

\section{Coherent and dissipative couplings facilitated by photon continuum}
     \label{Appx:PhBath}
 
Let us consider an array of QEs and metal nanoparticles shown in Fig.~\ref{Fig:PC}~(a). Each metal nanoparticle is assigned local surface plasmon (LSP) mode giving rise to the collective SPCM. Contribution of the dark plasmon mode is neglected in this Appendix. Further adopting the dipolar gauge~\cite{Cohen-TannoudjiPH:2014}, we treat interactions between the array components via quantized transverse electric field (i.e., transverse photons). The Hamiltonian of such a system is
\begin{eqnarray}
\label{H-net}
 	\hat{H} =\hat H_\text{ph}+\hat H_\text{sp}+\hat H_\text{qe},
\end{eqnarray}
where the first term stands for the non-interacting photon modes 
\begin{eqnarray}
\label{Hph-def}
\hat H_\text{ph} = \sum_{\bm q,\lambda=1,2}\omega_{\bm q}\hat a_{\bm q\lambda}^\dag \hat a_{\bm q\lambda},
\end{eqnarray}
with $\hat a_{\bm q\lambda}^\dag$  being a creation operator for a mode with the wave vector $\bm q$ and state of polarization $\lambda$. The photon frequency is $\omega_q=c|\bm q|$ with $c$ being speed of light.  

The second and the third terms in the Hamiltonian~(\ref{H-net}) 
\begin{eqnarray}
\label{HLSP}
\hat H_\text{sp} &=& \sum\limits_{\bar m}\hbar\omega_{\bar m}\hat\psi_{\bar m}^\dag\hat\psi_{\bar m} 
-\sum_{\bar m}\hat{\bm p}_{\bar m}\cdot \hat{\bm E}_{\bar m}^\bot,
\\\label{HQE}
\hat H_\text{qe} &=& \sum_{n}\hbar\omega_n\left(\hat s_{n}^z+\frac{1}{2}\right) 
-	\sum_{n}\hat {\bm\mu}_n\cdot \hat{\bm E}^\bot_n,\;\;\;\;
\end{eqnarray}
stand for the LSPs localized at sites $\bar m$ having energy $\hbar\omega_{\bar m}$ and described by the Bose operators $\{\hat\psi_{\bar m},\hat\psi_{\bar m}^\dag \}$ and the QEs with energies $\hbar\omega_n$ described by the site spin operators $\hat s^\alpha_{n}$, $\alpha = z, \pm$, respectively. The LSPs and QEs are coupled via transition dipole operators 
\begin{eqnarray}
\label{psp-odef}
\hat{\bm p}_{\bar m} &=& {\bm p}_{\bar m}(\hat\psi_{\bar m}^\dag+\hat\psi_{\bar m} ),
\\\label{muqe-odef}
\hat{\bm \mu}_n &=& \bm\mu_n\left(\hat s^+_n+\hat s^-_n\right),
\end{eqnarray}
to the  electric field $\hat{\bm E}_{\bar m}^\bot\equiv\hat{\bm E}^\bot(\bm r_{\bar m})$ and $\hat{\bm E}_n^\bot\equiv\hat{\bm E}^\bot(\bm r_n)$ with ${\bm p}_{\bar m}$ and ${\bm \mu}_n$ being the LSP and QE transition dipole matrix elements. The transverse electric field operator is
\begin{eqnarray}
\label{Eph-def}
	\hat{\bm E}^\bot(\bm r)&=&i\sum_{\bm q, \lambda=1,2}{\cal E}_{\bm q\lambda}
	\left(\hat a_{\bm q\lambda} e^{i\bm q\cdot\bm r}-\hat a^\dag_{\bm q\lambda} e^{-i\bm q\cdot\bm r}\right),
\end{eqnarray}
where the amplitude 
\begin{eqnarray}
\label{Eam-def}
{\cal E}_{\bm q\lambda} = \left(\frac{\hbar\omega_{q}}{2\varepsilon_o V}\right)^{1/2}\bm e_{\bm q\lambda},
\end{eqnarray}
depends on the mode polarization vector $\bm e_{\bm q\lambda}$, quantization volume $V$, and vacuum permittivity $\varepsilon_o$.  

The Heisenberg equations of motion for the LSP and QE operators due to the Hamiltonian~(\ref{H-net})--(\ref{HQE}) are
\begin{eqnarray}
\label{eqm-psi-def}
\partial_t\hat{\psi}_{\bar m} &=& -i\omega_{\bar m}\hat{\psi}_{\bar m} +\frac{i}{\hbar}\bm p_{\bar m}\cdot\hat{\bm E}_{\bar m}^\bot, 
\\\label{eqm-s-def}
\partial_t{\hat s_n^-}&=&  -i\omega_n \hat s_n^- 
						- \frac{2i}{\hbar}:\bm \mu_n\cdot\hat{\bm E}^\bot_n\hat s_n^z: ,
\\\label{eqm-sz-def}
\partial_t{\hat s_n^z} &=& 	-\frac{i}{\hbar}:\bm \mu_n\cdot\hat{\bm E}^\bot_n(\hat s_n^- - \hat s_n^+):,\;\;\;\;\;\;
\end{eqnarray}
where $:\hat A \hat B:$ denotes operators' $\hat A$ and $\hat B$ normal ordering. 

By integrating the photon field operator equation of motion
\begin{eqnarray}
\label{pht-eq-mov}
	\partial_t\hat a_{\bm q\lambda} &=& -i\omega_{\bm q}\hat a_{\bm q\lambda}
	+\frac{1}{\hbar}\sum_{\bar{m}}\hat{\bm p}_{\bar{m}}\cdot{\cal E}_{\bm q\lambda}e^{-i\bm q\cdot\bm r_{\bar m}} 
\\\nonumber&~&
	+\frac{1}{\hbar}\sum_n\hat{\bm \mu}_{n}\cdot{\cal E}_{\bm q\lambda}e^{-i\bm q\cdot\bm r_n},
\end{eqnarray}
and substituting the result into Eq.~\eqref{Eph-def}, one partitions the electric field operator at  site $\alpha=\{\bar m, n\}$ into three components~\cite{Milonni_QV:1994}
\begin{eqnarray}
\label{E-sol}	
	 \hat {\bm E}^\bot_{\alpha}(t) &=& \hat {\bm E}^\bot_{\text{in},\alpha}
	 +\hat {\bm E}^\bot_{\text{s},\alpha}+\hat {\bm E}^\bot_{\text{rr},\alpha}.
\end{eqnarray}
The first term, 
\begin{eqnarray}
\label{Ein-def}
\hat{\bm E}^\bot_{\text{in},\alpha}(t)&=& i\sum_{\bm q, \lambda=1,2}{\cal E}_{\bm q\lambda}
	\left(\hat a_{\bm q\lambda}(t_{o}) e^{- i\omega_q(t-t_o)+i\bm q\cdot\bm r_\alpha}\;\;\;
\right.\\\nonumber &~& \left.	
	-\hat a^\dag_{\bm q\lambda}(t_{o}) e^{i\omega_q(t-t_o)-i\bm q\cdot\bm r_\alpha}\right),	
\end{eqnarray}
is the input field due to the photon field initial condition at time $t_o$ propagated to time $t$ by the interaction free photon Hamiltonian~(\ref{Hph-def}). We identify this field as the vacuum fluctuations. 

The second term in Eq.~\eqref{E-sol} describes the field produced by dipoles at their own positions
\begin{eqnarray}
\label{ESP-rr-der}	
	 \hat {\bm E}^\bot_{\text{rr},\bar m} &=& \lim\limits_{r_{\bar m}\rightarrow 0}
	 \dd{\bf G}(r_{\bar m })\cdot\hat{\bm p}_{\bar m}(t-r_{\bar m}/c), 
\\\label{EQE-s-der}
	\hat {\bm E}^\bot_{\text{rr},n} &=&\lim\limits_{r_{n}\rightarrow 0}
	\dd{\bf G}(r_{n })\cdot\hat{\bm \mu}_{n}(t-r_{n}/c),
\end{eqnarray}
known as the radiation reaction field. The third term in Eq.~\eqref{E-sol} describes the scattered field 
\begin{eqnarray}
\label{ESP-s-der}	
	 \hat {\bm E}^\bot_{\text{s},\bar m} &=& \sum_{\bar m'\neq \bar m}
	 \dd{\bf G}_{\bar m \bar m'}\cdot\hat{\bm p}_{\bar m'}(t-r_{\bar m \bar m'}/c) 
\\\nonumber &+&	
	\sum_n \dd{\bf G}_{\bar m n}\cdot\hat{\bm\mu}_{n}(t-r_{\bar m n}/c), 
\\\label{EQE-s-der}
	\hat {\bm E}^\bot_{\text{s},n} &=&\sum_{\bar m}\dd{\bf G}_{n \bar m}\cdot\hat{\bm p}_{\bar m}(t-r_{n \bar m}/c),
\end{eqnarray}
due to the polarization of the QE and LSP dipoles surrounding either the $\bar m$-th LSP or $n$-th QE. Here, $r_{\alpha\beta}=|\bm r_\alpha-\bm r_{\beta}|$ and $\alpha,\beta=\{\bar m,n\}$. Notice that Eq.~\eqref{EQE-s-der}, neglects the contribution of the QEs and as a result the QE-QE interactions, justified by, $p_{\bar m}\gg\mu_n$. Finally, the photon time-domain dyadic Green function $\dd{\bf G}_{\alpha\beta}=\dd{\bf G}(r_{\alpha\beta})$ entering Eqs.~\eqref{ESP-rr-der}--\eqref{EQE-s-der} is
\begin{eqnarray}
\label{Grph-def}
\dd{\bf G}(\bm r)&=&
\frac{1}{4\pi\varepsilon_0}
\left\{\left({\bf n}\otimes{\bf n}-\dd{\bf I}\right)\frac{\partial_t^2}{c^2r}
\right.\\\nonumber &~&\left.
+\left(3{\bf n}\otimes{\bf n}-\dd{\bf I}\right)\left(\frac{\partial_t }{cr^2}
	+\frac{1}{r^3}\right)\right\},
\end{eqnarray}
where $\dd{\bf I}$ is the unit $3\times 3$ matrix and ${\bf n}=\bm r/r$ is a directional unit vector. 

Next, we use Eqs.~(\ref{eqm-psi-def}) and (\ref{eqm-s-def}) to eliminate the time derivatives in the right-hand-side of Eqs.~\eqref{ESP-rr-der}--\eqref{EQE-s-der} following by the short-time operator expansions $\hat\psi_{\bar m'}(t-r_{\alpha \bar m'}/c)\approx\hat\psi_{\bar m'}(t)e^{i\omega_{\bar m'}r_{\alpha \bar m'}/c}$ and $\hat s^-_{n'}(t-r_{\alpha n'}/c)\approx\hat s^-_{n'}(t)e^{i\omega_{n'}r_{\alpha n'}/c}$. This results in the radiation reaction terms describing the spontaneous radiative  decay of the LSPs and QEs
\begin{eqnarray}
\label{pErr-sp}
\frac{i}{\hbar}\bm p_{\bar m}\cdot\hat{\bm E}^\bot_{\text{rr},\bar m}&=& 
			-\frac{\kappa_{\bar m}}{2}(\hat{\psi}_{\bar m}-\hat{\psi}_{\bar m}^\dag), 						
\\\label{muErr-qe}
\frac{i}{\hbar}\bm \mu_n\cdot\hat{\bm E}^\bot_{\text{rr},n} &=& -\frac{\kappa_n}{2}
	:\left(1+ \hat\Pi_n\right)(\hat s_n^- -\hat s_n^+):, 			 	
\end{eqnarray}
with associated rates 
\begin{eqnarray}
\label{kappa-sp}
	\kappa_{\bar m}&=& \frac{p_{\bar m}^2\omega_{\bar m}^3}{3\pi\varepsilon_o\hbar c^3},
\\\label{kappa-qe}
	\kappa_n&=& \frac{\mu_n^2\omega_n^3}{3\pi\varepsilon_o\hbar c^3},
\end{eqnarray}
respectively. According to Eq.~\eqref{muErr-qe}, the energy transfer from QEs to LSP enhances the QE dephasing/decay rate as indicated by the polarization operator
\begin{eqnarray}\label{Pi_n}
\hat\Pi_n=\sum_{\bar m}\left(\frac{2\lambda_{n \bar m}}{\omega_n}\right)^2\left(2\hat\psi^\dag_{\bar m}\hat\psi_{\bar m}+1\right).
\end{eqnarray}

Respectively, evaluation of the scattered field for the LSPs and QEs gives 
\begin{widetext} 
\begin{eqnarray}
\label{Es-sp-rhs}
\frac{i}{\hbar}\bm p_{\bar m}\cdot\hat{\bm E}^\bot_{\text{s},\bar m}&=& 
		-i\sum_{\bar m'\neq\bar m} \lambda_{\bar m \bar m'}(\hat\psi_{\bar m'} + \hat\psi_{\bar m'}^\dag)
		-\sum_{\bar m'\neq\bar m} \eta_{\bar m \bar m'}(\hat\psi_{\bar m'} - \hat\psi_{\bar m'}^\dag)
\\\nonumber&~&			
			-i\sum_{n} \lambda_{\bar m n}(\hat s_n^- + \hat s_n^+)	
			-\sum_{n} \eta_{\bar m n}(\hat s_n^- - \hat s_n^+), 			 
\\\label{Es-qe-rhs}
\frac{i}{\hbar}\bm \mu_n\cdot\hat{\bm E}^\bot_{\text{s},n} &=& 
			-i\sum_{\bar m} \lambda_{n \bar m}(\hat\psi_{\bar m} + \hat\psi_{\bar m}^\dag)
		-\sum_{\bar m} \eta_{n \bar m}(\hat\psi_{\bar m} - \hat\psi_{\bar m}^\dag).
\end{eqnarray}
\end{widetext}
Here, the {\em coherent}, $\lambda_{\alpha\beta}$, and {\em dissipative}, $\eta_{\alpha\beta}$, coupling parameters naturally appear after elimination of the photon degrees of freedom and have the following representation
\begin{widetext}
\begin{eqnarray}
\label{lambda_ab}
	\lambda_{\alpha\beta}&=&\frac{3d_\alpha}{4d_\beta}\kappa_\beta \left[
	-\xi_{\alpha\beta}\frac{\cos(q_\beta r_{\alpha\beta})}{q_\beta r_{\alpha\beta}}
	+\zeta_{\alpha\beta}\left( 
		\frac{\sin(q_\beta r_{\alpha\beta})}{(q_\beta r_{\alpha\beta})^2}
		+\frac{\cos(q_\beta r_{\alpha\beta})}{(q_\beta r_{\alpha\beta})^3}
	\right)	\right],
\\\label{eta_ab}
	\eta_{\alpha\beta}&=&\frac{3d_\alpha}{4d_\beta}\kappa_\beta  \left[\;\;
	\xi_{\alpha\beta}\frac{\sin(q_\beta r_{\alpha\beta})}{q_\beta r_{\alpha\beta}}
	+\zeta_{\alpha\beta}\left(
		\frac{\cos(q_\beta r_{\alpha\beta})}{(q_\beta r_{\alpha\beta})^2}
		-\frac{\sin(q_\beta r_{\alpha\beta})}{(q_\beta r_{\alpha\beta})^3}
	\right)	\right],
\end{eqnarray}
\end{widetext}
where 
\begin{eqnarray}
\label{ksi_ab}
	\xi_{\alpha\beta}&=&\bm{\mathfrak{n}}_\alpha\cdot\bm{\mathfrak{n}}_{\beta}
		-({\bf n}_{\alpha\beta}\cdot\bm{\mathfrak{n}}_\alpha)({\bf n}_{\alpha\beta}\cdot\bm{\mathfrak{n}}_\beta),
\\\label{zeta_ab}
	\zeta_{\alpha\beta}&=&\bm{\mathfrak{n}}_\alpha\cdot\bm{\mathfrak{n}}_{\beta}
		-3({\bf n}_{\alpha\beta}\cdot\bm{\mathfrak{n}}_\alpha)({\bf n}_{\alpha\beta}\cdot\bm{\mathfrak{n}}_\beta),		
\end{eqnarray}
and $\bm{\mathfrak{n}}_{\alpha}=\bm d_{\alpha}/d_{\alpha}$, $\bm{\mathfrak{n}}_{\beta}=\bm d_{\beta}/d_{\beta}$, ${\bf n}_{\alpha\beta}=\bm r_{\alpha \beta}/r_{\alpha\beta}$, $q_\beta=\omega_\beta/c$, and $r_{\alpha\beta}=|\bm r_\alpha-\bm r_\beta|$ with $\alpha,\beta=\{\bar m, n\}$; $\bm d_{\bar m} = \bm p_{\bar m}$, $\bm d_{n} = \bm\mu_n$. The spontaneous decay rates $\kappa_\beta$, $\beta=\{\bar m,n\}$ are given in Eqs.~\eqref{kappa-sp} and \eqref{kappa-qe}.

The substitution of Eq.~\eqref{E-sol} along with Eqs.~(\ref{pErr-sp}), (\ref{muErr-qe}), (\ref{Es-sp-rhs}), and (\ref{Es-qe-rhs}) into Eqs.~(\ref{eqm-psi-def})--(\ref{eqm-sz-def}) results in a set of quantum Langevin equations for coupled LSP and QE operators  
\begin{widetext}
\begin{eqnarray}
\label{eqm-psi-full}
\partial_t {\hat\psi}_{\bar m} &=& -i\omega_{\bar m}\hat{\psi}_{\bar m} 
	-\frac{\kappa_{\bar m}}{2}\left({\hat\psi}_{\bar m}-{\hat\psi}_{\bar m}^\dag\right)
	-i\sum_{\bar m'\neq\bar m} \lambda_{\bar m \bar m'}(\hat\psi_{\bar m'} + \hat\psi_{\bar m'}^\dag)
	-\sum_{\bar m'\neq\bar m}\eta_{\bar m \bar m'}(\hat\psi_{\bar m'}-\hat\psi_{\bar m'}^\dag)
\\\nonumber&~&
	-i\sum_{n} \lambda_{\bar m n}(\hat s_n^- + \hat s_n^+) 
	-\sum_{n} \eta_{\bar m n}( \hat s_n^- -  \hat s_n^+) 
	+\frac{i}{\hbar}\bm p_{\bar m}\cdot\hat{\bm E}^\bot_{\text{in},\bar m},
\\\label{eqm-s-full}
\partial_t{ \hat s_n^-}&=&-i\omega_n \hat s_n^- 
	-\frac{\kappa_n}{2}:\left(1+ \hat\Pi_n\right)(\hat s_n^- - \hat s_n^+):
	+2i\sum_{\bar m} \lambda_{n \bar m}:\hat s_n^z(\hat\psi_{\bar m} + \hat\psi_{\bar m}^\dag):
\\\nonumber&~&
 + 2\sum_{\bar m}\eta_{n \bar m}:\hat s_n^z(\hat\psi_{\bar m} - \hat\psi_{\bar m}^\dag):
 -\frac{2i}{\hbar}:\bm \mu_n\cdot\hat{\bm E}^\bot_{\text{in},n}\hat s_n^z:,	
\\\label{eqm-sz-av}
\partial_t{ \hat s_n^z} &=& - \kappa_n\left(1+ \hat\Pi_n\right)\left( \hat s_n^z+\frac{1}{2}\right)
		+i\sum_{\bar m} \lambda_{n \bar m}:(\hat s_n^- - \hat s_n^+)(\hat\psi_{\bar m} + \hat\psi_{\bar m}^\dag):
\\\nonumber	&~&
		+\sum_{\bar m}\eta_{n \bar m}:( \hat s_n^- - \hat s_n^+)(\hat\psi_{\bar m} - \hat\psi_{\bar m}^\dag):
		-\frac{i}{\hbar}:\bm \mu_n\cdot\hat{\bm E}^\bot_{\text{in},n}(\hat s_n^- - \hat s_n^+):.		
\end{eqnarray}
\end{widetext}
According to Eqs.~\eqref{eqm-psi-full}--\eqref{eqm-sz-av}, the photon continuum fluctuations (i.e., $\hat{\bm E}^\bot_{\text{in}}$) result in cooperative dissipation processes such as spontaneous radiative decay of LSPs including the superradiant emission due to the LSP-LSP coupling with the rate $\eta_{\bar m \bar m'}$, the QE radiative decay, and the LSP-QE dissipative coupling with the rate $\eta_{n \bar m}$.    

The Dicke model assumes identical LSP (QE), frequencies $\omega_{\bar m}=\omega_{sp}$ ($\omega_{n}=\omega_{o}$) and transition dipoles ${\bm p}_{\bar m}={\bm p}$  (${\bm\mu}_{n}={\bm \mu}$) corresponding to identical spontaneous radiative decay rates $\kappa_{\bar m}=\kappa_{sp}$ ($\kappa_{n}=\kappa_o$). Furthermore, we introduce collective LSP modes characterized by momentum $k$  
\begin{eqnarray}
\label{psi_k}
	\hat\psi_k=\frac{1}{\sqrt{\bar{\cal N}}}\sum\limits_{\bar m=1}^{\bar{\cal N}} \hat\psi_{\bar m} e^{-ik{\bar m}} ,		
\end{eqnarray}
where $\bar{\cal N}$ is the number of metal nanoparticles in the array. Identifying the SPCM as the $k=0$ collective mode, $\hat\psi\equiv\hat\psi_{k=0}$, we simplify Eqs.~\eqref{eqm-psi-full}--\eqref{eqm-sz-av} to the following form
\begin{widetext}
\begin{eqnarray}
\label{eqm-psi}
\partial_t {\hat\psi} &=& -i(\omega_{sp}+\lambda_{sp}){\hat\psi}-\frac{1}{2}(\kappa_{sp}+\eta_{sp})(\hat\psi-\hat\psi^\dag)
  -i\lambda\sum_{n} \left(\hat s_n^- +\hat s_n^+\right) -\eta\sum_{n} \left(\hat s_n^- -\hat s_n^+\right)
  +\frac{i}{\hbar}\bm p\cdot\hat{\bm E}^\bot_{\text{in},sp},
\\\label{eqm-sn-}
\partial_t{ \hat s_n^-}&=&-i\omega_o\hat s_n^{-} 
		- \frac{1}{2}\left[\kappa_o + \gamma_{sp}\hat\psi^\dag\hat\psi \right](\hat s_n^{-}-\hat s_n^{+})  
		 +2i\lambda :\hat s_n^z(\hat\psi + \hat\psi^\dag):+ 2\eta:\hat s_n^z(\hat\psi - \hat\psi^\dag):
		 -\frac{2i}{\hbar}:\bm \mu\cdot\hat{\bm E}^\bot_{\text{in},n}\hat s_n^z:,		
\\\label{eqm-sz}
\partial_t{ \hat s_n^z} &=& - \left[\kappa_o+ \gamma_{sp}\hat\psi^\dag\hat\psi\right]\left(\hat s_n^z+\frac{1}{2}\right)
+i\lambda :\left(\hat s_n^- -\hat s_n^+\right) (\hat\psi + \hat\psi^\dag):	
+ \eta :\left(\hat s_n^- -\hat s_n^+\right)(\hat\psi - \hat\psi^\dag):
\\\nonumber&~&
-\frac{i}{\hbar}:\bm \mu\cdot\hat{\bm E}^\bot_{\text{in},n}(\hat s_n^- - \hat s_n^+):,	
\end{eqnarray}
\end{widetext}
where 
\begin{eqnarray}
\label{Einsp}
\hat{\bm E}^\bot_{\text{in},sp} &=& \frac{1}{\sqrt{\bar{\cal N}}}\sum\limits_{\bar m=1}^{\bar{\cal N}}
	\hat{\bm E}^\bot_{\text{in},\bar m}. 
\end{eqnarray}
In Eqs.~\eqref{eqm-psi}--\eqref{eqm-sz}, the SPCM band edge renormalization,  superradiant decay, and the QE plasmon assisted spontaneous decay rates are   
\begin{eqnarray}
\label{lmbd_sp}
\lambda_{sp}&=&\frac{1}{\bar{\cal N}}\sum\limits_{\bar m=1}^{\bar{\cal N}-1}\lambda_{\bar m\bar{\cal N}}, 
\\\label{eta_sp}
\eta_{sp}&=&\frac{2}{\bar{\cal N}}\sum\limits_{\bar m=1}^{\bar{\cal N}-1}\eta_{\bar m\bar{\cal N}}, 
\\\label{g_sp}
\gamma_{sp}&=&2\bar{\cal N}\kappa_{sp}\left(\frac{\lambda}{\omega_o}\right)^2,
\end{eqnarray}
respectively. Respectively, the SPCM-QE coherent, $\lambda$, and dissipative, $\eta$ coupling rates are
\begin{eqnarray}
\label{lmbd_spqe}
\lambda &=&\frac{1}{\sqrt{\bar{\cal N}}}\sum\limits_{\bar m=1}^{\bar{\cal N}}\lambda_{n\bar m}, 
\\\label{eta_spqe}
\eta&=&\frac{1}{\sqrt{\bar{\cal N}}}\sum\limits_{\bar m=1}^{\bar{\cal N}}\eta_{n\bar m}.
\end{eqnarray}
Above, we also assumed that the QEs are arranged so that each has identical coherent and dissipative coupling rates to the SPCM.  

The model represented by Eqs.~\eqref{eqm-psi}--\eqref{eta_spqe} recovers the generalized Dicke model introduced in Sec.~\ref{Sec:SPDicke}, provided the counter-rotating dissipative terms and the plasmon assisted QE decay operator (i.e., $\gamma_{sp}\hat\psi^\dag\hat\psi$) are dropped and the SPCM and QE nonradiative decay and dephasing rates are added.  

 
 \section{Two-mode surface plasmon model allowing for dissipative coupling}
 \label{Appx:2SPM}

The Dicke Hamiltonian accounting for two surface plasmon modes coupled to the QEs and with each other reads
\begin{eqnarray}
 \label{Hpl-2SPM}
 \hat H &=& \Omega\hat\phi^\dag\hat\phi + \omega_o\hat\psi^\dag\hat\psi 
 		+\omega_o\left(\sum\limits_n\hat s_n^z+\frac{{\cal N}_o}{2}\right) 
 \\\nonumber &~&
 		+\zeta\hat Q\hat q +2\left(\xi\hat Q+ \lambda\hat q\right)
 		\sum\limits_n\hat s_n^{x}.
 \end{eqnarray}
Here and below, the quadrature representation $\hat Q = \hat\phi + \hat\phi^\dag$, $\hat q =\hat \psi + \hat\psi^\dag$, $\hat s_n^x=(s_n^{+} +\hat s^{-}_n)/2$, and $\hat s_n^y= i(s_n^{-} -\hat s^{+}_n)/2$ is used for the sake of brevity. The coupling parameters $\zeta$, $\xi$, and $\lambda$ can be obtained microscopically in the dipole-dipole approximation using an approach developed in Ref.~\cite{ZasterJPCS:2016}. 

 A set of the Heisenberg equations of motion due to the Hamiltonian~\eqref{Hpl-2SPM} reads
 \begin{eqnarray}
 \label{Heqm-phi}
 \partial_t {\hat\phi} &=& -(i\Omega +\Lambda){\hat\phi}
     -i\zeta\hat q-2i\xi\sum\limits_n\hat s_n^x, 
 \\\label{Heqm-psi}
 \partial_t {\hat\psi} &=& -(i\omega_o +\Gamma_{sp}){\hat\psi}
     -i\zeta\hat Q -2i\lambda \sum\limits_n\hat s_n^x, 
 \\\label{Heqm-sn-}
 \partial_t\hat s_n^{-}&=&-\left(i\omega_o+\Gamma_o\right)\hat s_n^{-}
 	+2i\hat s_n^z \left(\xi\hat Q + \lambda\hat q \right),	
 \\\label{Heqm-snz}
 \partial_t\hat s_n^z &=& -\gamma_o(\hat s_n^z-d_o/2)
                 + 2\hat s_n^y\left(\xi\hat Q + \lambda\hat q\right),		
 \end{eqnarray}
where the SPCM dephasing rate $\Gamma_{sp}$, the QE dephasing, $\Gamma_o=(\gamma_\downarrow/2+\gamma_\uparrow/2+\gamma_\phi)$ and population decay $\gamma_o=\gamma_\downarrow+\gamma_\uparrow$ rates, and the inversion parameter $d_o=(\gamma_\uparrow-\gamma_\downarrow)/(\gamma_\uparrow+\gamma_\downarrow)$ are added as discussed in Sec.~\ref{Sec:SPDicke}.

 Integrating Eq.~\eqref{Heqm-phi} in the Markovian approximation, and neglecting the counter-rotating terms containing $e^{(\omega_o+\Omega)t'}$ gives
 \begin{eqnarray}
 \label{phi(t)-Mark}
 \hat\phi(t) &=& \hat \phi(0)e^{-(i\Omega +\Lambda)t}
 \\\nonumber &~&   
   - \frac{\Delta+i\Lambda}{\Delta^2 +\Lambda^2} 
     \left[\zeta\hat \psi(t)+\xi\sum\limits_n\hat s_n^{-}(t)\right], 
 \end{eqnarray}
with the detuning $\Delta = \Omega-\omega_o$. Taking into account that the dark plasmon mode is broad, $\Delta\ll\Lambda$, we can safely set $\Delta=0$ in Eq.~\eqref{phi(t)-Mark} and neglect contribution of the initial condition $\hat \phi(0)$ which should decay at time $t$.  

By making substitution of Eq.~\eqref{phi(t)-Mark} with $\Delta=0$ and $\hat \phi(0)=0$ into Eqs.~\eqref{Heqm-psi}--\eqref{Heqm-snz}, we obtain our final set of operator equations where the dark plasmon mode is integrated out
\begin{widetext}
 \begin{eqnarray}
 \label{Heqm-psi-f}
 \partial_t {\hat\psi} &=& -(i\omega_o +\gamma_{sp}){\hat\psi}-\eta_{sp}(\hat\psi-\hat\psi^\dag)
      -i\lambda\sum\limits_{n}\left(\hat s_n^{-} +\hat s_n^{+}\right)
      -i\eta\sum\limits_{n}\left(\hat s_n^{-} -\hat s_n^{+}\right), 
 \\\label{Heqm-sn-f}
 \partial_t\hat s_n^{-}&=&-\left(i\omega_o+\gamma_o\right)\hat s_n^{-}
 	-\eta_{qe}(\hat s_n^{-}-\hat s_n^{+}) +2\eta_{qe}:s_n^z\sum\limits_{m\neq n}\left(\hat s_m^{-}-\hat s_m^{+}\right):
	+i\lambda:\hat s_n^z (\hat\psi +\hat\psi^\dag):
	-\eta:\hat s_n^z \left(\hat\psi -\hat\psi^\dag\right):,	
\\\label{Heqm-snz-f}
 \partial_t\hat s_n^z &=& -\gamma_t\left(\hat s_n^z-\frac{d_o}{2}\right)
 				-2\eta_{qe}\left(\hat s_n^z+\frac{1}{2}\right)
				+\eta_{qe}:\left(\hat s_n^{-}-\hat s_n^{+}\right)\sum\limits_{m\neq n}\left(\hat s_m^{-}-\hat s_m^{+}\right):
                + i\lambda :\left(\hat s_m^{-}-\hat s_m^{+}\right)(\hat\psi+\psi^\dag):
\\\nonumber &~&                
              +:\left(\hat s_n^{-}-\hat s_n^{+}\right)(\hat\psi+\psi^\dag):  .		
 \end{eqnarray}
\end{widetext}
Here $:\hat A \hat B:$ denotes operators' $\hat A$ and $\hat B$ normal ordering. Due to the darks plasmon mode Eqs.~\eqref{Heqm-psi-f}--\eqref{Heqm-snz-f} aquire the dissipative rates $\eta_{sp}=\zeta^2/\Lambda$,   $\eta_{qe}=\xi^2/\Lambda$, and $\eta=\xi\zeta/\Lambda$, describing the SPCM decay, QEs cooperative decay, and the SPCM-QE dissipative coupling, respectively. 

The generalized Dicke model in Sec.~\ref{Sec:SPDicke} can be recovered from Eqs.~\eqref{Heqm-psi-f}--\eqref{Heqm-snz-f} by neglecting the terms $2\eta_{qe}:s_n^z\sum_{m\neq n}\left(\hat s_m^{-}-\hat s_m^{+}\right):$ and $\eta_{qe}:\left(\hat s_n^{-}-\hat s_n^{+}\right)\sum_{m\neq n}\left(\hat s_m^{-}-\hat s_m^{+}\right):$ and subsequently dropping the counter-rotating dissipative terms.

\section{Generalized Tavis-Cummings model}
\label{Appx:GTC}

We generalize a driven-dissipative Tavis-Cummings model by including the dissipative coupling between the SPCM and QEs. Associated equations of motion 
\begin{eqnarray}
\label{TC-psi}
\partial_\tau \psi &=& -\left(i+\bar\Gamma_{sp}\right)\psi
 -\bar\lambda (i+\bar\eta){\cal N}_o s_{-}, 
\\\label{TC-s-}
\partial_\tau s_{-}&=&-\left(i+\bar\Gamma_o\right)s_{-}
	+2\bar\lambda (i+\bar\eta)s_z\psi,	
\\\label{TC-sz}
\partial_\tau s_z &=& -\bar\gamma_o(s_z-d_o/2)
	+2\bar\lambda\text{Re}\left[(i-\bar\eta)\psi^* s_{-}\right],		
\end{eqnarray}
result from Eqs.~\eqref{mf-SPcoh}-\eqref{mf-spop} after applying the rotating wave approximation. 
 
We further break the $U(1)$ gauge symmetry of Eqs.~\eqref{TC-psi}--\eqref{TC-sz} by fixing the phase $\psi\rightarrow e^{i\bar\omega_l\tau}\psi $, $s_-\rightarrow e^{i\bar\omega_l\tau} s_-$ with $\bar\omega_l=\omega_l/\omega_o$ to be identified as the normalized lasing frequency. As a result, the equations of motion for the coherences become 
\begin{eqnarray}
\label{TC-psi-1}
\partial_\tau \psi &=& -\left(i\delta\bar\omega_l+\bar\Gamma_{sp}\right)\psi
 -\bar\lambda (i+\bar\eta){\cal N}_o s_{-}, 
\\\label{TC-s-1}
\partial_\tau s_{-}&=&-\left(i\delta\bar\omega_l+\bar\Gamma_o\right)s_{-}
	+2\bar\lambda (i+\bar\eta)s_z\psi,			
\end{eqnarray}
where $\delta\bar\omega_l=1-\bar\omega_l$ denotes a normalized frequency detuning. Eq.~\eqref{TC-sz} does not change under such a transformation. Using the steady state solution of Eq.~\eqref{TC-s-1} 
\begin{eqnarray}
\label{s-st}
s_- = \frac{2\bar\lambda(i+\bar\eta)}{i\delta\bar\omega_l+\bar\Gamma_o}s_z\psi,	
\end{eqnarray}
we eliminate $s_{-}$ in Eq.~\eqref{TC-psi-1} and in the steady state solution of Eq.~\eqref{TC-sz} to obtain 
\begin{eqnarray}
\label{TC-psi-2}
\partial_\tau \psi &=& -\left[\left(i\delta\bar\omega_l+\bar\Gamma_{sp}\right)
 +2\bar\lambda^2{\cal N}_o \frac{(i+\bar\eta)^2}{i\delta\bar\omega_l+\bar\Gamma_o} s_z\right]\psi, 
\\\label{TC-sz-2}
s_z &=& \frac{d_o}{2}\left[1+4\bar\lambda^2\frac{\bar\Gamma_o}{\bar\gamma_o}\frac{1+\bar\eta^2}{\delta\bar\omega_l^2+\bar\Gamma_o^2}|\psi|^2\right]^{-1}.	
\end{eqnarray}

The lasing threshold can be found by equating to zero both the real and imaginary parts of the expression in the square brackets of Eq.~\eqref{TC-psi-2}. Roots of these two equation with respect to $\delta\bar\omega_l$ and $s_z$ stand for the lasing frequency detuning, $\delta\bar\omega_l(\bar\eta)$, Eq.~\eqref{dwl-eta}, and the expression for the steady state population inversion   
\begin{eqnarray}
\label{sz-c}
s_z = \frac{d_o}{2}\frac{\bar\lambda_l^2(\bar\eta)}{{\cal N}_o\bar\lambda^2}.	
\end{eqnarray}
Here, the critical coupling $\bar\lambda_l^2(\bar\eta)$ is defined by Eq.~\eqref{lmbc-laser} and the expression is valid for $\bar\lambda_l^2(\bar\eta)\leq \bar\lambda^2$. Finally,  making the substitution of Eq.~\eqref{sz-c} into Eq.~\eqref{TC-sz-2}, one finds the order parameter,  
\begin{eqnarray}
\label{psi-c}
|\psi|^2 &=& \frac{\bar\gamma_o(\delta\bar\omega^2_l(\bar\eta)+\bar\Gamma_o^2)}
 {4\bar\Gamma_o\bar\lambda_l^2(\bar\eta)(1+\bar\eta^2)}
 \left(1-\frac{\bar\lambda_l^2(\bar\eta)}{{\cal N}_o\bar\lambda^2}\right),	
\end{eqnarray}
representing the SPCM spontaneous coherence for $\bar\lambda^2\geq\bar\lambda_l^2(\bar\eta)$. Expression for the QE spontaneous coherence 
\begin{eqnarray}
\label{s-c}
|s_-|^2 = \frac{\bar\gamma_o\bar\lambda_l^2(\bar\eta)}{\bar\Gamma_o{\cal N}_o\bar\lambda^2}
 \left(1-\frac{\bar\lambda_l^2(\bar\eta)}{{\cal N}_o\bar\lambda^2}\right),	
\end{eqnarray}
follows from Eqs.~\eqref{s-st} and \eqref{psi-c}  


\section{Relationship between output electric field and photon emission energy spectrum}
\label{Appx:InOut}

Taking into account that the surface plasmon transition dipole exceeds the QE ones, i.e., $p_{sp}\gg\mu$, the cavity emission is mostly contributed by the SPCM which indeed can be strongly coupled to the QEs. To find an  output electric field produced by the cavity in the far-field photo-detector detector zone, we employ the Heisenberg equation of motion~\eqref{pht-eq-mov} for the photon mode operator in which the right-hand-side  QE term is dropped by setting $\mu_n=0$.  

Following the input output formalism~\cite{Gardiner_IO:1985,Gardiner_Qnoise:2004}, we integrate Eq.~\eqref{pht-eq-mov} both forward ($t>t_o$) and backward ($t_f>t$) in time. This provides us with the expression  for the cavity output field
\begin{eqnarray}
\label{E_in_out-G}
	&~&\hat {\bm E}_{\tt out}^\bot(\bm r, t) = \hat {\bm E}^\bot_\text{in}(\bm r, t) 
\\\nonumber	&~&\;\;\;\;\;\;\;\;
	+\sum\limits_{\bar m}\dd{\bf G}(\bm r-\bm r_{\bar m})\cdot\hat{\bm p}_{\bar m}(t-|\bm r-\bm r_{\bar m}|/c),
\end{eqnarray}
in terms of the input filed (Eq.~\eqref{Ein-def}) and the LSP transition dipole operator $\hat{\bm p}_{\bar m}$ (Eq.~\eqref{psp-odef}) propagated by the dyadic photon Green function (Eq.~(\ref{Grph-def})) to the detector at coordinate $\bm r$. 

Since, the cavity-detector separation significantly exceeds the cavity linear size and $t\gg |\bm r-\bm r_{\bar m}|/c$, we can replace the LSP mode with the collective SPCM (Eq.~\eqref{psi_k}) and retain only the far-field, term in the photon Green function. Finally, partitioning the output field into the negative and positive frequency components, $\hat {\bm E}_{\tt out}^\bot=\hat{\bm E}^{-}_{\tt out}+\hat{\bm E}^{+}_{\tt out}$,~\cite{Milonni-PRA:1995}, we obtain 
\begin{eqnarray}
\label{EFF-in-out+}
	&~&\hat {\bm E}_{\tt out}^{-}(\bm r,t) = \hat {\bm E}^{-}_\text{in}(\bm r, t)
	-\frac{{\bm p}_{sp}-{\bf n}({\bf n}\cdot{\bm p}_{sp})}{4\pi\varepsilon_0c^2 r}
	\partial_t^2 \hat\psi^\dag(t),\;\;\;
\\\label{EFF-in-out-}
	&~&\hat {\bm E}_{\tt out}^{+}(\bm r,t) = \hat {\bm E}^{+}_\text{in}(\bm r, t)
	-\frac{{\bm p}_{sp}-{\bf n}({\bf n}\cdot{\bm p}_{sp})}{4\pi\varepsilon_0c^2 r}
	\partial_t^2 \hat\psi(t),
\end{eqnarray}
where $\bm r$ denotes the radius vector from the cavity and a detector, ${\bf n} = \bm r/r$ with $r=|\bm r|$, and $\hat{\bm E}^{-}_\text{in}=(\hat{\bm E}^{+}_\text{in})^\dag $ is due to the first term in the right-hand-side of Eq.~\eqref{Ein-def}.

The power emitted by the cavity is 
\begin{eqnarray}
\label{Ispt} 
		\frac{dE}{dt}=2\varepsilon_o c r^2 \int\limits_{4\pi^2}d\Omega\left\langle \hat{\bm E}^{-}_{\tt out}(\bm r,t)
			\cdot\hat{\bm E}^{+}_{\tt out}(\bm r,t) \right\rangle,
\end{eqnarray}
where the integration is perform over the whole solid angle surrounding the cavity. The substitution of Eqs.~\eqref{EFF-in-out+} and \eqref{EFF-in-out-} into Eq.~\eqref{Ispt} and subsequent integration over the solid angle results in
\begin{eqnarray}
\label{Ispt-1} 
		\frac{dE(t)}{dt}=\frac{p_{sp}^2}{3\pi \varepsilon_o c^3} 
		\left\langle \partial_t^2\hat\psi^\dag(t)\partial_t^2\hat\psi(t) \right\rangle.
\end{eqnarray}
The correlation functions containing the input field fluctuations vanish after averaging. 
 
Following the Wiener-Khintchine theorem~\cite{Milonni_QO:2019}, we define the photon power spectrum in terms of the SPCM operator auto-correlation function as 
\begin{eqnarray}
\label{S-WK} 
	S(\omega)\delta(\omega-\omega')&=&
	\int_{-\infty}^\infty \frac{dt}{2\pi}\int_{-\infty}^\infty \frac{dt'}{2\pi}
\\\nonumber&\times&	
	\langle\partial_t^2\hat\psi^\dag(t)\partial_t^2\hat\psi(t')\rangle
	e^{i\omega t-i\omega't},
\end{eqnarray}
which  simplifies to the form
\begin{eqnarray}
\label{S-WK-psi} 
	S(\omega)\delta(\omega-\omega')&=&\omega^2 \omega^{'2} 
	\int_{-\infty}^\infty \frac{dt}{2\pi}\int_{-\infty}^\infty \frac{dt'}{2\pi}
\\\nonumber&\times&	
	\langle\hat \psi^\dag(t)\hat\psi(t')\rangle
	e^{i\omega t-i\omega't'}.
\end{eqnarray}
For the stationary process, one integral in the right-hand side of Eq.~\eqref{S-WK-psi} can be evaluated resulting in $\delta(\bar\omega-\bar\omega')$ to recover the form of Eq.~\eqref{Sw} for $S(\omega)$. According to Eqs.~\eqref{Ispt-1}--\eqref{S-WK-psi}, the photon energy emitted by the cavity per frequency range $d\omega$ is related to the spectral function $S(\omega)$ as
\begin{eqnarray}
\label{dE/dw} 
		\frac{dE(\omega)}{d\omega}=\frac{p_{sp}^2}{3\pi \varepsilon_o c^3} S(\omega).
\end{eqnarray}
Note that an expression for the (power) spectrum of the photons emitted  within infinitesimally small solid angle $d\Omega$, can be obtained by simply multiplying (Eq.~\eqref{Ispt-1}) Eq.~\eqref{dE/dw} with the prefactor $3 \sin^2\theta/(8\pi)$ where the angle $\theta$ is measured between $\bm p_{sp}$ and the radius vector pointing to the detector.

%
\end{document}